\documentstyle[prl,aps]{revtex}
\input epsf

\tighten
\begin{document}

\newcommand{\be}{\begin{equation}} 
\newcommand{\ee}{\end{equation}}


\title{Zero modes of fermions with general mass matrix}

\author{
Glenn Starkman$^1$, Dejan Stojkovic$^2$ and Tanmay Vachaspati$^{1}$}

\address
{${}^1$Department of Physics,
Case Western Reserve University,
10900 Euclid Avenue, Cleveland, OH 44106-7079, USA\\
\smallskip
${}^2$Department of Physics,
University of Alberta,
Edmonton, AB T6G 2J1, Canada}

\wideabs{
\maketitle

\begin{abstract}
\widetext

We analyze zero energy solutions of the Dirac equation
in the background of a string-like configuration in an extension 
of the standard model which accommodates the most general fermionic
mass matrix for neutrinos. If either the left- or the right-handed 
Majorana mass vanishes, neutrino and electron zero modes are found 
to exist. If the harmonic mean of the neutrino Majorana masses is 
large compared to the Dirac mass, we can prove that normalizable 
neutrino zero modes cease to exist. This leads to an odd number of 
fermionic zero modes -- namely, only the electron zero mode -- which
we argue implies that the bosonic background is in a topologically
distinct sector from the vacuum. This is confirmed by noting that 
the bosonic sector of the model involves the breaking of a global 
$U(1)$ symmetry and hence possesses topological global $U(1)$ strings.
\end{abstract}
}

\narrowtext

It is well appreciated that in bosonic field theories whose
vacuum manifold is topologically non-trivial, classical, static
solutions of the field equations can exist and be stable. These
solutions globally minimize the field energy in the topological
sector in which they lie. The fermionic modes were examined
in the background of the topological solutions and often one or
more zero modes are found. Using certain ``index theorems'', the
zero modes can often be directly related to the topological
properties of the background. In this sense, the zero modes
are an outcome of the background topology.

The bosonic sector of the minimal electroweak standard model
does not have the non-trivial topology required for the existence
of topologically stable strings. However, there exist non-topological,
string-like configurations of Higgs and gauge fields, usually called
Z-strings \cite{AchVac00} which are solutions of the bosonic equations 
of motion. Fermionic zero modes on a Z-string background have already
been constructed \cite{EarPer94,GarVac95}. However, since this 
construction has
been in the framework of the standard model, the fermions only have
Dirac masses and, in particular, the neutrino is massless. We would
like to examine the zero modes in an extension of the electroweak
standard model in which fermions have both Dirac and Majorana masses,
that is, a general mass matrix.
We will adopt the standard notation for the various fields
({\it eg.} see \cite{AchVac00,StaStoVac01}). In
addition to the usual standard model fields, we will include a complex
Higgs field $\Delta$, transforming in the adjoint representation
of $SU(2)$. The full Lagrangian is:

\begin{equation} \label{L}
{\cal L} = {\cal L}_{EW} + {\cal L}_{\Delta} + {\cal L}_{f}
\label{fullL}
\end{equation}
where,
\begin{equation}
{\cal L}_{EW}=
 -{1\over 4} W^a_{\mu\nu}W^{a\mu\nu}-
         {1\over 4} F_{\mu\nu}F^{\mu\nu}+
                 |D_\mu\Phi |^2 + V(\Phi )
\label{LEW}
\end{equation}
\begin{equation}
{\cal L}_{\Delta} =
|D_\mu \Delta |^2 + U(\Delta , \Phi )
\end{equation}
\begin{eqnarray}
{\cal L}_{f} = &&
i{\bar\Psi}\gamma^\mu D_\mu\Psi +
           i\overline{e_R}\gamma^\mu D_\mu e_R +
             i\overline{\nu_R}\gamma^\mu \partial_\mu \nu_R +
\nonumber \\
&-& [ h'\overline{e_R}\Phi^\dagger\Psi
+h\overline{\nu_R}\left(\Phi^T (i\tau_2)^{\dagger}\Psi\right)
 + \nonumber \\
&+& {1\over 2} h'' \bar{\Psi} i\tau_2 \Delta \Psi^c
+{1\over 2} \overline{\nu_R} M_R(\nu_R)^c + h.c.] \ .
\end{eqnarray}
\noindent $U(\Delta ,\Phi )$ and $V(\Phi )$ are general 
gauge invariant quartic potentials:
\be
V(\Phi )=- \lambda\left(\Phi^\dagger\Phi-\eta^2 \right)^2
\ee
\begin{eqnarray}
U(\Delta ,\Phi )= && a Tr(\Delta^\dagger \Delta) +  b 
(Tr(\Delta^\dagger \Delta))^2  \nonumber \\
&+& c \Phi^\dagger \Phi Tr(\Delta^\dagger \Delta) +
d Tr(\Delta^\dagger \Delta^\dagger) Tr(\Delta \Delta) \nonumber\\
&+& e \Phi^\dagger \Delta^\dagger \Delta \Phi +
(f \Phi^T \Delta i \tau_2 \Phi + h.c.) \ . 
\label{potentialU}
\end{eqnarray}
Parameters $a$, $b$, $c$, $d$, $e$, $f$, $\lambda$, $\eta$, $h$, $h'$ 
and $h''$ are constants.
 The covariant derivatives are defined as:
\begin{eqnarray}
D_\mu \Psi &=& \left(\partial_\mu-i{g\over2}\tau^a W^a_\mu
+i{{g'}\over2}Y_\mu\right)\Psi \nonumber \nonumber \\
D_\mu e_R &=& \left(\partial_\mu+ig'Y_\mu\right)e_R \\
D_\mu\Phi &=&  \left(\partial_\mu-i{g\over2}\tau^a W^a_\mu
-i{{g'}\over2}Y_\mu\right) \Phi \nonumber \\
D_\mu \Delta &=& \partial_\mu \Delta +i g'Y_\mu \Delta
-i{g\over2}  \left[\tau^a W_\mu^a, \Delta \right]  \ . \nonumber
\end{eqnarray}
where $g$ and $g'$ are the gauge coupling constants while
$\tau^a$ are the Pauli matrices .
The fields in the model are defined as:
$\Psi=\left( \nu_L  , e_L  \right)^T$, $\nu^c = C \bar\nu^T$ and
$\Psi^c = C \bar\Psi^T$ where $C \equiv i \gamma^2 \gamma^0$ is the
charge conjugation matrix.
$W^a_{\mu\nu}$ and $F_{\mu\nu}$ are the field strength tensors
for $W_\mu^a$ and $Y_\mu$ respectively.
The Higgs triplet $\Delta$ transforms according to the adjoint 
representation of $SU(2)_L$ and with hypercharge $-2$ can be
written as 
\begin{equation}
 \Delta = \pmatrix{H^+& H^{++} \cr
                   H^0 & -H^+} \ .
\end{equation}
The gauge symmetry of the Lagrangian (\ref{L}) is $[SU(2)_L \times 
U(1)_Y]/Z_2$.
If the coefficient of the last term in the potential $U(\Delta, \Phi)$
is zero, {\it i.e.} $f=0$, then  the bosonic sector of the Lagrangian 
(\ref{L}) possesses an additional global $U(1)$ symmetry
\be
\Delta \rightarrow \Delta ' = e^{i \alpha} \Delta \ , \ \ \ \ \
\phi \rightarrow \phi ' = \phi \ ,
\ee
which we call $U(1)_L$. This symmetry provides lepton number 
conservation
in the bosonic sector of the theory. Lepton number conservation in the
fermionic sector is explicitely violated by the presence right-handed 
Majorana mass, $M_R$, for neutrinos. However, if $M_R$ were to arise
from the vacuum expectation value of another scalar field with 
non-trivial
lepton number, $U(1)_L$ could also be implemented as a symmetry of the
fermionic sector.
From now on we set the coefficient  $f$ to zero by requiring the 
explicit
$U(1)_L$ symmetry in the bosonic potential before the phase transition.

The parameters of the model can be chosen so that $\Delta$ and
$\Phi$ get vacuum expectation values:
\begin{equation}
h'' \Delta = \pmatrix{0 & 0 \cr M_L & 0} , \ \
\Phi = \pmatrix{0 \cr \eta}
\end{equation}
with $M_L$ and $\eta$ being constants.
This breaks the original $[SU(2)_L \times U(1)_Y]/Z_2$ gauge symmetry 
down to
$U(1)_{EM}$. Constraining analysis to the gauge symmetries (the 
important effects of global symmetries will be appropriately analysed later)  
we find that
the vacuum manifold is an $S^3$ and contains no incontractable
paths. This means that there are no topological string solutions in the
model. However, there do exist non-topological Z-string
solutions \cite{AchVac00} and these shall form the basis of the 
background bosonic configuration that we shall study.

Let us define the $Z$ gauge field by
\begin{equation}
Z_\mu \equiv \cos\theta_w W^3_\mu-\sin\theta_w Y_\mu
\end{equation}
where $\theta_w$ is the weak mixing angle ($\tan \theta_w \equiv 
g'/g$).
Then the bosonic field configuration that we choose to study is:
\begin{equation}
qZ_\theta = - {{v(r)}\over r}\ ,
\ \ \Phi = \phi \pmatrix{0\cr 1}\ , \ \
h'' \Delta = \pmatrix{0 & 0 \cr M_L & 0}
\label{background}
\end{equation}
with
\begin{equation}
\phi \equiv \eta f(r)e^{i\theta}
\label{littlephi}
\end{equation}
and all other gauge fields set to zero. Also,
$q \equiv \sqrt{g^2+g'^2} /2$, $v(r)$ and $f(r)$ are
profile functions, and $(r,\theta ,z)$ are cylindrical coordinates.
Note that the configuration is independent of the $z$ coordinate.

Without the $\Delta$ field, the field equations of motion can
be solved and this determines the profile functions $f(r)$ and
$v(r)$. The solution is known as the $Z$-string.
It is unstable except for large $\theta_w$ and
small values of the ratio of scalar ($\Phi$) to $Z$ mass.
With the $\Delta$ included, the configuration (with $f$ and
$v$ as in the $Z$-string) is not a solution.  This can easily be seen
since $D_\theta \Delta \propto Z_\theta \Delta \propto 1/r$ even as
$r \rightarrow \infty$ while the potentials $U$ and $V$ go to zero
exponentially fast at large $r$. The energy of the configuration is
logarithmically divergent, reminiscent of the global $U(1)$ string.
None of these features really concerns us since we are free to
consider fermionic modes on any bosonic background we wish. The
fermionic zero modes that we will obtain will be well-localized.

The neutrino and electron equations of motion in the bosonic
background are:
\begin{eqnarray} \label{DiraceqnsN}
i\gamma^\mu D_\mu \nu_L &=& h\phi^\ast \nu_R + M_L (\nu_L)^c 
\nonumber\\
i\gamma^\mu \partial_\mu \nu_R &=& h\phi \nu_L + M_R (\nu_R)^c \\
i\gamma^\mu D_\mu e_L &=& h'\phi e_R \nonumber \\
i\gamma^\mu D_\mu e_R &=& h'\phi^\ast e_L  \ .
\label{Diraceqns}
\end{eqnarray}

In the standard model, {\it i.e.} in the absence of $\Delta$, $M_L$
and $M_R$, these Dirac equations have been solved. The fermionic
equations are known to have a well-localized ({\it i.e.} with an
exponential fall off) zero mode just as in the case of
topological vortices \cite{JacRos81}. There is a zero mode
for the electron propagating in one direction along the string
and another zero mode for the neutrino propagating in the opposite
direction \cite{EarPer94,GarVac95}.
In the case of massless neutrinos, {\it i.e.} $h\eta=M_L=M_R=0$,
the zero mode is not discretely normalizable but is delta function
normalizable. In other words, it is part of a continuum
of modes \cite{StaStoVac01}. If we include a massive neutrino with
$h \eta \neq 0$, $M_R \neq 0$ and $M_L = 0$ there is one
well-localized, discretely normalizable solution. The electron
equations in (\ref{Diraceqns}) also have one well-localized
discrete zero mode regardless of the values of $M_L$ and $M_R$.

\begin{figure}[tbp]
\centerline{\epsfxsize = 0.60 \hsize \epsfbox{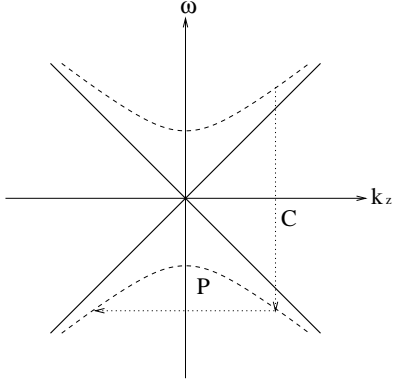}}
\caption{\label{fig1}
In the usual case, the fermionic spectrum contains two lines
($\omega=\pm k_z$)
that describe zero modes boosted in the $z$-direction. One line
describes the left-moving and the other the right-moving zero modes.
Perturbations of the bosonic background can mix
the two zero modes at the origin and convert them into two massive
modes that lie on hyperbolae. The points on the two hyperbolae
are related by CP transformations as shown.}
\end{figure}

Since the $Z$-string configuration is non-topological, it is
interesting to study the fate of the zero modes when the background
is perturbed. It has been shown that the perturbations in the Higgs and
gauge sectors continuously mix the electron and neutrino zero modes
(see Fig. \ref{fig1}) and convert the zero modes into two low lying
massive states \cite{Nac95,LiuVac96}. In this process, the number of
eigenmodes is conserved, only the initially vanishing eigenvalues
have flowed to non-vanishing values.

Now we study neutrinos with the general mass matrix:
\begin{equation}
\label{mm} \cal{M}=\pmatrix{M_L & M_D \cr M_D & M_R}
\label{massmatrix}
\end{equation}
where
the Dirac mass $M_D = h\eta f(r) e^{-i \theta}$.
The electron still has a zero mode since its equations are
decoupled from those of the neutrino.
Now we shall
prove that if $M_L$ and $M_R$ are large enough
compared to $M_D$, then there is no neutrino zero mode on
the string.

Let us write the neutrino Dirac equations as:
\begin{equation}
i\gamma^\mu {\bf D}_\mu \Psi = {\bf M} \Psi
\label{Diracequation}
\end{equation}
where
\begin{equation}
\Psi = \pmatrix{\nu_L \cr \nu^c_R \cr \nu^c_L \cr \nu_R \cr } \ ,
\ \
{\bf M} = \pmatrix{0&0&M_L&M_D \cr
                   0&0&M_D&M_R \cr
                   M^\dagger_L&M^\dagger_D&0&0 \cr
                   M^\dagger_D&M^\dagger_R&0&0 }
\label{Psi}
\end{equation}
and ${\bf D}_\mu$ is matrix valued so as to recover
the original Dirac equations (\ref{DiraceqnsN}). We can get some
information about the spectrum of the Dirac operator by squaring
it. If we take $\Psi ({\vec x} ,t) = e^{i\omega t} \psi ({\vec x})$
with the normalization $\int d^2x \, \psi^\dagger \psi = 1$,
and act by $i\gamma^\mu {\bf D}_\mu$ on eq. (\ref{Diracequation}),
we get
\begin{eqnarray}
\omega^2 =&& - \int d^2x\, \psi^\dagger {\bf D}^2 \psi +
\int d^2x\, \psi^\dagger [\sigma^{\nu \mu}F_{\mu \nu} + |M_D|^2]\psi
                               \nonumber \\
&&
+4 \int d^2x \, (T[\alpha ,\delta ]+T[\beta , \gamma ] )
\nonumber
\end{eqnarray}
where the contribution of the Majorana masses is all in the
function
$T$:
\begin{eqnarray}
T[\alpha , \delta ] \equiv &&
|M_L|^2 |\alpha |^2 +
|M_R|^2 |\delta |^2 +
\nonumber \\ &&
(M_L M_D^\dagger + M_D M_R^\dagger ) \alpha \delta +
\nonumber \\ &&
(M_L^\dagger M_D + M_D^\dagger M_R ) \alpha^* \delta^*
\label{defineT}
\end{eqnarray}
and
\begin{equation}
\nu_L = \pmatrix{\alpha\cr \beta\cr -\alpha\cr -\beta\cr} \ ,
\ \
\nu_R = \pmatrix{\gamma\cr \delta\cr \gamma\cr \delta\cr} \ .
\label{components}
\end{equation}
If the Majorana masses vanish, then one sets $T =0$. Here we
know for sure that there is a zero mode {\it i.e.} a choice of
$\psi$ such that $\omega^2 =0$. So let us write:
\begin{equation}
\omega^2 = \Omega[\psi ] +
        4 \int d^2x (T[\alpha ,\delta ] + T[\beta , \gamma ])
\end{equation}
>From the zero Majorana mass case, we know that the minimum value
of the functional $\Omega$ is zero. That is, no matter what $\psi$
we choose, we cannot make $\Omega$ negative. The least we can
make it is zero. Then if we can prove that $T$ is positive
(and non-vanishing) for all $\psi$, we will have shown that
$\omega^2 > 0$ and that there is no neutrino zero mode.

It is not hard to see that $T[\alpha ,\delta ] > 0$ for any
choice of $\alpha$ and $\delta$ provided
\begin{equation}
|M_L M_R| >
 {\rm max}_{(r,\theta )}|M_L M_D^\dagger + M_D M_R^\dagger |
\label{condition}
\end{equation}
Note that $M_D$ is a function of $(r,\theta )$ and we
would like to find the maximum value of the expression
on the right-hand side. The inequality can be simplified
further by using the form of $M_D$ given below eq. (\ref{massmatrix}).
Then we find that the condition for $T > 0$ is:
\begin{equation}
{1 \over {|h| \eta}} > {1\over {|M_L|}} +  {1\over {|M_R|}} 
\label{etacondition}
\end{equation}
that is, the harmonic mean of the Majorana masses must be larger
than $|h| \eta$.
One can check that if this condition is saturated, then it is
possible to find $\alpha$ and $\delta$ such that $T=0$ at some
spatial points.
Therefore the condition in eq. (\ref{condition}) is equivalent
to having $T[\alpha,\delta]>0$ for all $\alpha,\delta,r$ and $\theta$.
Hence if the parameters satisfy equation (\ref{etacondition}), then
there are no normalizable neutrino zero modes\footnote{
A similar example of fermionic zero modes on
domain walls where the existence of zero modes depends on 
the values of the mass matrix parameters may be found in 
\cite{Sto01}.}.
We expect (but have not shown) that if the parameter $M_R$ 
(for example) is varied from $M_R =0$ to a large value satisfying 
the condition in eq.  (\ref{etacondition}), the neutrino zero mode 
solution that exists for small $M_R$ will change and cease to become 
normalizable at some critical $M_R$. 

\begin{figure}[tbp]
\centerline{\epsfxsize = 0.60 \hsize \epsfbox{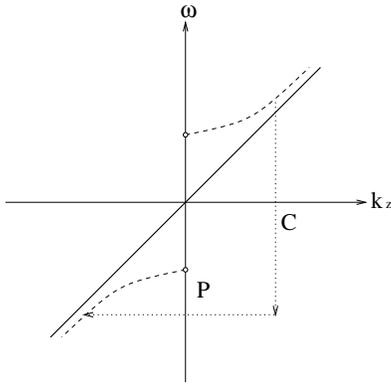}}
\caption{ \label{fig2}
In the present case, the spectrum contains only the line of boosted
electron zero modes ($\omega= +k_z$). Perturbations of the background 
cannot change the line into the spectrum for massive fermions 
(hyperbolae). 
In particular, perturbations cannot convert the single zero mode
($\omega =0=k_z$) into a massive mode because CP invariance would 
require it to split into two modes, leading to a discontinuous 
spectrum.}
\end{figure}

The possibility of not having a neutrino zero mode while the
corresponding electron zero mode is still present is important
from the point of view of the string stability under
perturbations. In this case, perturbations of the
string cannot lift the electron zero mode into a massive mode.
A single zero mode (see Fig. \ref{fig2}) cannot be
continuously deformed as in the case described by Fig. \ref{fig1}.
The central point $(\omega , k) =(0,0)$ cannot be moved up or down
without violating CP invariance\footnote{
It is easy to check that the bosonic background in
eq. (\ref{background}) remains invariant under CP transformations.}.
Yet there is no electron zero mode in the vacuum. For non-topological
strings this conclusion seems
paradoxical because there is no topology in the bosonic sector
that prevents the string from decaying into the vacuum.

To understand the non-trivial topology arising due to the fermionic 
zero modes, note that we have set $f=0$ in the bosonic potential
in eq. (\ref{potentialU}). The absence of the corresponding
terms means that we can globally rotate $\Phi$ and $\Delta$
by a relative phase and this will not change the potential
energy. Hence, the bosonic sector has an extra global $U(1)$
symmetry that is broken once $\Phi$ and $\Delta$ acquire
vacuum expectation values. Therefore there is indeed a
broken global $U(1)$ (i.e. the $U(1)_L$) in the model and this means 
that
global $U(1)$ strings can be present. The winding of the
relative phase of $\Delta$ and $\Phi$ must be non-trivial
around such a string and is precisely given by the winding
of the operator $\Phi^T \Delta i \tau_2 \Phi$ (last term
of the potential in eq. (\ref{potentialU})). 

Alternately we could study fermionic zero modes on our choice
of background when the parameter $f$ in the potential does not 
vanish. In this case, the global $U(1)$ symmetry is absent from
the model and, from the odd number of fermionic zero modes, it 
might seem that one still has a topological index arising due
to the fermions (see Fig. 2). However, in this case, the last
term in the potential (\ref{potentialU}) is non-trivial and
contributes to the energy density. In fact, it does not vanish
at spatial infinity and gives rise to an energy that diverges
in proportion to the integration volume. In such a situation,
there is no reason to require the fermionic zero modes to be
well-localized and indeed to be normalizable -- the fermions
will be affected by the energy density at infinity. A proper
interpretation of the number of zero modes is only possible
when the asymptotic bosonic field configuration is in its
vacuum.

In conclusion, we have studied fermionic zero modes on certain
string-like bosonic backgrounds when the fermions have a
general mass matrix. We come to the interesting conclusion 
that the number of fermionic zero modes is a function of the 
fermionic mass parameters. In our model, for small harmonic
mean of the Majorana masses, the difference of the number of 
normalizable left-moving and right-moving fermion zero modes 
($|n_L-n_R|$) vanishes, while for large harmonic mean of the
Majorana masses, $|n_L-n_R| =1$. The underlying model --- a trivial 
extension
of the standard model with massive neutrinos --- supports topologically 
stable
``electroweak" strings.

\acknowledgements

TV is grateful to Roman Jackiw, Tom Kibble, Ashoke Sen and Diptiman Sen 
for comments.  This work was supported by the DoE to the
Department of Physics, CWRU and the Killam Trust, UoA to DS.

\end{document}